# The text2term tool to map free-text descriptions of biomedical terms to ontologies

Rafael S. Gonçalves[1]*, Jason Payne[1], Amelia Tan[2], Carmen Benitez[3], Jamie Haddock[3], Robert Gentleman[1]

[1] Center for Computational Biomedicine, Harvard Medical School, Boston, MA, USA
[2] Department of Biomedical Informatics, Harvard Medical School, Boston, MA, USA
[3] Harvey Mudd College, Los Angeles, CA, USA

**\*** Corresponding author (rafael_goncalves@hms.harvard.edu)

## Abstract

There is an ongoing need for scalable tools to aid researchers in both retrospective and prospective standardization of discrete entity types—such as disease names, cell types or chemicals—that are used in metadata associated with biomedical data. When metadata are not well-structured or precise, the associated data are harder to find and are often burdensome to reuse, analyze or integrate with other datasets due to the upfront curation effort required to make the data usable—typically through retrospective standardization and cleaning of the (meta)data. With the goal of facilitating the task of standardizing metadata—either in bulk or in a one-by-one fashion; for example, to support auto-completion of biomedical entities in forms—we have developed an open-source tool called text2term that maps free-text descriptions of biomedical entities to controlled terms in ontologies. The tool is highly configurable and can be used in multiple ways that cater to different users and expertise levels—it is available on PyPI and can be used programmatically as any Python package; it can also be used via a command-line interface; or via our hosted, graphical user interface-based Web application (https://text2term.hms.harvard.edu); or by deploying a local instance of our interactive application using Docker.

**Database URL:** https://pypi.org/project/text2term and https://text2term.hms.harvard.edu

## Introduction

Scientific data repositories typically require data contributors to create metadata that describe their datasets, as a way to enhance the discoverability, usefulness and overall quality of the data. However, there is often insufficient control of the metadata associated with scientific datasets when that data is deposited in data repositories, resulting in poor quality metadata that hinder the useability of those datasets. For instance, in a study of the quality of metadata in the NCBI BioSample [1]—a repository of metadata about biological samples—the authors found substantial variability in the values given by users for key database fields such as *disease [2]*, where data authors provide syntactically different strings to denote the same disease or condition (e.g., "cardiac failure", "heart failure", "myocardial failure"). This is the kind of problem that the many controlled vocabularies, medical terminologies and ontologies are designed to mitigate—for example, the Unified Medical Language System (UMLS) [3] contains 185 vocabularies and ontologies used in the medical domain; the BioPortal [4] ontology repository contains over 1,100 biomedical ontologies. As a result of these quality issues, data are harder to find and are often burdensome to reuse, analyze, or integrate with other datasets due to the upfront curation effort required to make the data usable—typically through retrospective standardization and cleaning of the (meta)data. The broader consequence to science is that the effectiveness of performing biomedical research through reuse of archived data is substantially diminished.

With growing awareness and adoption of the FAIR data principles [5] to make data Findable, Accessible, Interoperable, and Reusable, scientists are increasingly attempting to add semantic annotations to their data. Although there are various tools that enable users to annotate free-text in their metadata with one or more ontologies, it is unusual to see real-time control of metadata in data submission portals such as auto-complete-type input fields for key metadata elements describing the submitted data that use ontologies or controlled vocabularies as providers of acceptable input values. This could be in part because existing tools are primarily designed to facilitate the *retrospective* standardization of discrete entities, metadata elements, such as names of diseases, phenotypes, cell types or chemicals.

There are two broad mechanisms for mapping terms to ontologies: programmatically, by leveraging dedicated packages, command-line interfaces or APIs; or interactively, by leveraging user interface-based tools developed for the purpose. In this paper, we describe the *text2term* tool to facilitate both programmatic and interactive mapping (or *grounding*) of plain-text entity descriptions to controlled terms in ontologies. The tool is open-source, openly developed and distributed under an MIT license. A Python package—available in the Python Package Index (PyPI) repository of software for the Python programming language—provides the main engine for programmatic mapping. A React JavaScript-based Web application provides user interfaces for interactive mapping which can be deployed on any machine using Docker. The driving design rationale for the tool was to facilitate the annotation of metadata containing thousands of free-text descriptions of phenotypes [under analysis in Genome-Wide Association Studies (GWAS)] with ontologies. The end-goal was to enable integration with and search across biobank-scale resources, which are typically annotated with the Experimental Factor Ontology (EFO) [6], and large health insurance databases. EFO is used to annotate phenotypes in the GWAS Catalog database [7]—the NHGRI-EBI Catalog of human genome-wide association studies; in the OpenGWAS database [8]; and in the OpenTargets platform [9]. We evaluate the effectiveness of text2term based on how accurately it can re-find mappings that have been identified and accepted by human experts, and which are openly available to use and to download. Our benchmark test corpus consists of three mapping sets: the Biomappings [10] collection of community-curated and contributed ontology mappings; the ontology mappings hosted in the Ontology Lookup Service (OLS); and the mapping set of UK Biobank phenotypes to EFO [11]. We show that text2term selects the correct mappings with an accuracy of 79%, 81% and 73% when compared to benchmark mappings, respectively.

# Related Work

The BioPortal Annotator [12] is a Web service that takes a text input and returns a list of annotations, each consisting of an ontology term and its corresponding location in the text. It can be used through simple Web interfaces or via REST APIs. BioPortal Annotator uses a concept recognition system called Mgrep [13] to perform the string-matching between the text and the ontology terms. Mgrep is a fast and scalable string-matching tool that uses a tree-based data structure to store the ontology terms and their synonyms. Mgrep can handle exact and approximate matching, as well as regular expressions and wildcards.

CEDAR [14] develops a tool for metadata management—the CEDAR Workbench [15]—that can be configured to constrain inputs in specific (metadata) form fields to terms in one or more ontologies, or specific branches of ontologies. When a user is filling in a metadata form, CEDAR uses the BioPortal Annotator service to get matches between the user input and ontology terms.

Zooma [16] is a Web service that maps free text annotations to ontology terms based on a curated repository of annotation knowledge. It is designed to help researchers annotate their experimental data with ontology terms. The tool uses a combination of curated mappings in existing datasets and standard text matching to map text to ontologies.

SORTA [17] is a system for ontology-based re-coding and technical annotation of biomedical phenotype data. It maps input string values to a target ontology using Lucene and n-gram based matching. Lucene is an open-source text search engine library that allows fast and flexible indexing and querying of text data. SORTA is open-source and available as a Web service.

The clinical Text Analysis and Knowledge Extraction System (cTAKES) [18] is an open-source tool for mapping free-text descriptions of biomedical entities to ontology terms. It uses Apache's Unstructured Information Management Architecture framework and the OpenNLP toolkit to create linguistic and semantic annotations of clinical texts at scale.

MARIE [19] is an unsupervised learning-based tool designed to find controlled terms for input strings. MARIE employs a unique combination of string-matching methods and term embedding vectors generated by BioBERT. This approach allows the tool to utilize both structural and contextual information to calculate similarity measures between source and target terms.[1]

MetaMap [20] is a mapping tool developed by the National Library of Medicine that provides access to the concepts in the Unified Medical Language System (UMLS) Metathesaurus. The mapping process begins with lexical/syntactic analysis, which includes tokenization, sentence boundary determination, acronym/abbreviation identification, and part-of-speech tagging. MetaMap's string-matching algorithm links the processed text to the knowledge embedded in the Metathesaurus, including synonym-type relationships.

---

[1] The tool could not be executed at the time of writing and is not actively maintained.

Overall, interactive mapping is generally not well supported in the current landscape of tools. Some of the tools, such as SORTA or cTAKES, include user interfaces that present the mapping or annotation results, however none of them support editing the mappings in-place, nor do they provide in-place visual context of the mappings (e.g., the neighboring ontology terms of each mapping). Programmatic mapping is also not well supported, with only BioPortal and Zooma providing adequate interfaces for this purpose. However, the use of these services relies on internet access, and often the services impose measures to curb requests in bulk by throttling them, which results in slow response time when mapping many terms. Overall, there currently is no one solution that provides both easy programmatic operation and user interface support for visualizing, verifying, and editing the generated mappings.

## The *text2term* Tool

The text2term mapping tool (Figure 1) is designed to be a flexible computational instrument to generate potential ontology term mappings for arbitrary strings, using one of several similarity metrics and leveraging ontology details such as synonyms. It was built out of the need to have both programmatic and interactive mechanisms to compute and verify mappings of free-text descriptions of phenotypes in biomedical metadata to one or more OWL ontologies.

### Tool Design and Features

The text2term tool provides support for (a) popular edit distance metrics such as the Levenshtein distance commonly used in spell checkers; (b) an approach that is based on vectors computed using Term Frequency-Inverse Document Frequency (TF-IDF), which is a popular metric used in information retrieval; (c) an interface to the BioPortal Annotator tool, which allows mapping to any ontology in the BioPortal repository—the largest repository of biomedical ontologies; and (d) an interface to the Zooma Annotator tool, which allows mapping to any ontology in the Ontology Lookup Service (OLS) repository of biomedical ontologies. The text2term tool can be used in two ways: (1) *programmatically* by importing the text2term package from a Python environment, or by using a command-line interface; and (2) *interactively* via a Web application with user interfaces for input entry, mapping visualization, verification, and download. The Web application can be deployed locally using Docker, which is the ideal way to support bulk or repetitive mapping jobs; it can also be used directly on our hosted version at https://text2term.hms.harvard.edu, which is practical for smaller jobs.

Once installed, the text2term package is available to be imported from Python environments and used programmatically. The tool is also executable from the command-line, which does not require any programming. Multiple configuration options, described next, allow users to fine-tune text2term both programmatically and through the command line. The documentation of the tool is provided on GitHub both as a conventional README and a GitHub Page (https://ccb-hms.github.io/ontology-mapper), and in the *readthedocs* platform—a widely used, open-source software documentation hosting platform (https://text2term.readthedocs.io). Additional details about the tool are provided in the Supplementary Material, along with a complete list of features in Supplementary Table S1.

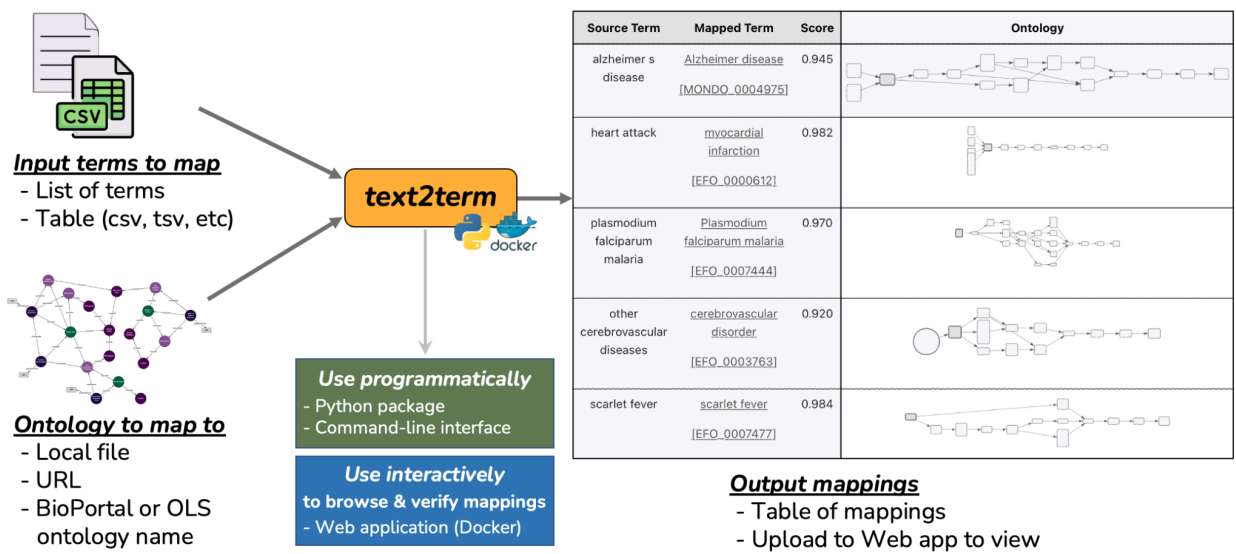

**Figure 1:** Overview of the text2term tool to map input terms (provided as a list of strings or as a table) to an ontology (specified by a file path, URL, or the ontology name as used by BioPortal or OLS for their Annotator and Zooma services).

**Input terms to map.** Text2term takes as input a list of strings, or a file containing strings to be mapped, which we call *source terms*. The file input can be either a text file containing a line-separated list of terms, or a table file where cells are delimited by some character (such as CSV or TSV) that can be configured via an argument in text2term.

**Ontology input and processing.** Text2term requires a *target ontology*, which can be any ontology specified in the W3C standard Web Ontology Language (OWL) format; the de-facto ontology language used nowadays. The target ontology can be provided as a local file path or as a URL to an ontology resource that can be resolved using standard URL handling methods—in our case, we use Python's built-in *urllib* URL handling modules. Furthermore, text2term also can work with ontologies hosted in the popular BioPortal and Ontology Lookup Service (OLS) repositories of biomedical ontologies.

We use the *owlready2 [21]* library to load ontologies, to perform ontology reasoning, and to obtain ontology term details such as labels, synonyms, and definitions. Specifically, the tool collects the following details for each ontology term:

1. Human-readable labels specified via *rdfs:label* or *skos:prefLabel* relationships.
2. Synonyms specified via *obo:hasExactSynonym* or via the ontology-specific synonym relationships in the NCI Thesaurus (*NCIT:P90*) and in the Experimental Factor Ontology (*EFO:alternative_term*). Optionally, broad synonyms (*obo:hasBroadSynonym*) can be included.
3. Definitions specified via skos:definition or the IAO definition property (*IAO:0000115*).

4. Hierarchical (*rdfs:SubClassOf*) relationships between ontology terms; specifically, we collect the parents, children, and instances of each ontology term. This can be done before or after reasoning, depending on the user configuration—by default, reasoning is not performed.

The ontology term details collected in steps 1-3 are used to match the input strings with ontology terms, while currently the details from step 4 are used for visualization purposes.

**Matching methods.** The text2term tool is designed to support, in a pluggable way, multiple string-matching techniques that users can select from. We mostly reuse existing methods and implementations available for string comparison. However, because of the poor performance we initially observed using these methods, we decided to design our own, simple approach that converts strings to vectors (using TF-IDF) and operates based on vector (cosine) similarity.

*TF-IDF-based mapper.* TF-IDF is a statistical measure often used in information retrieval and text mining—essentially, it measures how important a token is to a document relative to its occurrence across a corpus of documents. In our case, we consider as documents the terms in an ontology—specifically, the labels and exact synonyms of ontology terms. We obtain TF-IDF-based vectors for the input query strings and ontology terms, and then we use the widely used cosine similarity measure to compare vectors, which is the cosine of the angle between two vectors. In our implementation we use the *scikit-learn* library [23] to compute TF-IDF, and the *sparse-dot-topn* library [24] to compute cosine similarity between TF-IDF vectors. After experimenting with other implementations of cosine similarity—such as the built-in function in *scikit-learn*—we concluded that the *sparse-dot-topn* implementation provides a faster, more memory efficient way to compute the multiplication of two sparse matrices and to obtain the *top-n* closest values per query.

*BioPortal and Zooma Web API-based mappers*. We implemented Python interfaces to obtain mappings from two popular Web services for annotating unstructured text with ontology terms—the BioPortal Annotator and the Zooma annotator in OLS—through their respective Web/REST APIs. Our BioPortal mapping interface enables programmatically annotating terms with any ontology in the BioPortal repository by specifying the ontology name as used in BioPortal. Similarly, our Zooma interface allows annotating terms with ontologies available in the OLS repository by using their respective names as used in OLS. Ontology names are often, but not always the same between the two repositories. BioPortal does not return confidence scores for its annotations. We used a mapping score of 1 for all BioPortal annotations.

*Syntactic distance-based mappers*. We also provide support for commonly used and well known syntactic (edit) distance metrics. Specifically, we implemented support for matching input strings with ontology terms using the Levenshtein, Jaro, Jaro-Winkler, Jaccard, and Indel metrics.

The syntactic distance-based mappers as well as the Web API-based mappers perform slowly, since they do pairwise comparisons between each input string and each ontology term label and/or synonym, and there are networking and API load overheads for the Web API-based approaches. The TF-IDF-based approach reduces the mapping problem to matrix operations that are much faster to compute—this makes it possible to process tens of thousands of input queries in less than one minute.

## Interactive Web Application

We developed a Web application with user interfaces designed to facilitate the process of specifying tool inputs and to enable interactive curation of the mappings generated by text2term. The frontend of the Web application is implemented using React—a JavaScript library for building user interfaces. In the backend we use the Flask framework to implement RESTful APIs that allow interfacing with the text2term package, resuming a mapping session, or obtaining mapping results. We provide a Docker image that can be deployed in any machine with two simple commands.

The front page of the application is shown in Figure 2, where users specify, at a minimum, the input *source* terms, and the *target* ontology to map the terms to. Some additional, basic options are included in the UI to allow users to configure the tool. It is possible to resume a mapping session by clicking on the *Resume Mapping* link and then uploading a previously downloaded mapping table. The tool will effectively display the mapping results table marked up just as it was upon downloading. After submitting source terms and target ontology for mapping, users are presented with a table of mapping results, as shown in Figure 3. The mapping results contain the input source terms, their top-ranked mapped terms in the ontology, and then a visualization (shown in more detail in Figure 4) of where in the ontology class hierarchy those mapped terms are located.

The base functionality for the interface is the same as that of the programmatic tool described above. However, it should be noted that there are several features and options in the programmatic tool that are not available on the interface. This includes (but is not limited to) features such as caching, term-type specification, and including unmapped terms in the output.

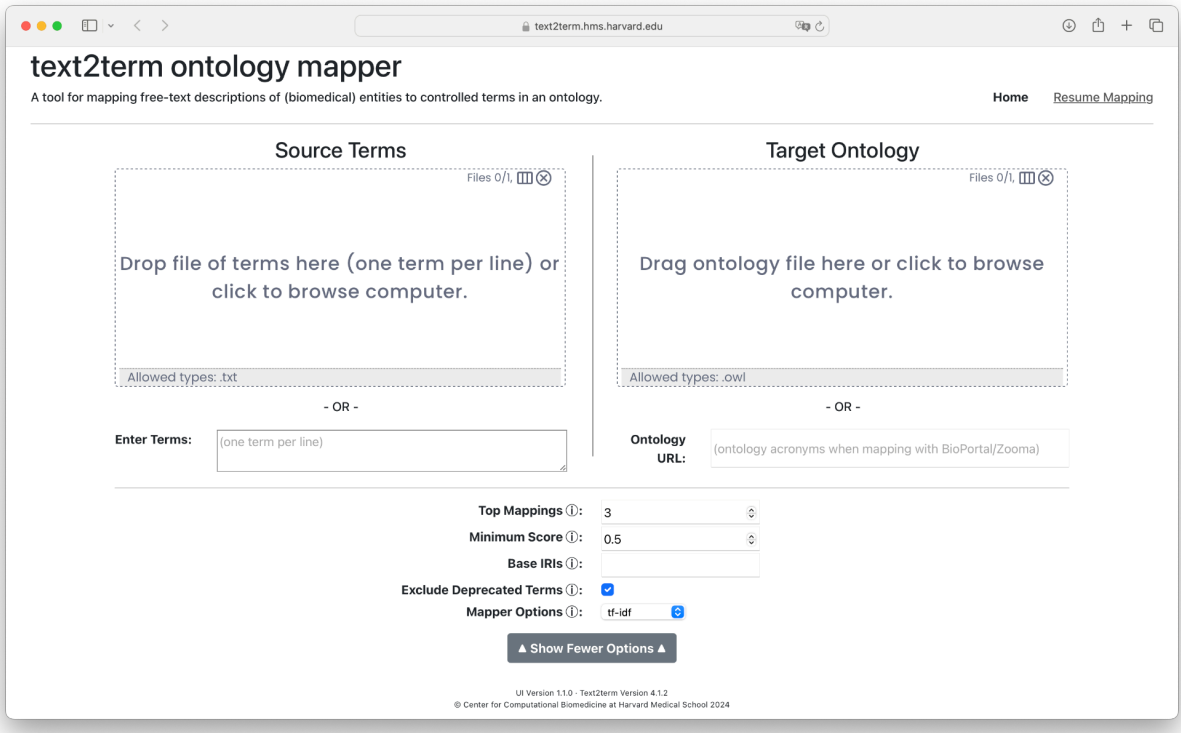

**Figure 2:** Front page of the text2term Web application (https://text2term.hms.harvard.edu). Here users specify input terms by uploading a file or entering raw text, and the target ontology to map those terms to, which can be uploaded or pointed to via its URL. Some additional, basic options are available in the UI, such as limiting the number of mappings per input term, specifying the minimum acceptable mapping score, the mapping method, etc.

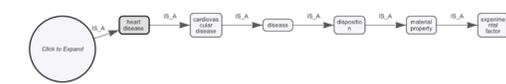

**Figure 3:** Results page showing mappings generated by text2term. From left to right, the table contains in each row: a button to *view alternate mappings* that shows mappings with lower

scores, which users can then choose from; the *Source Term* that was given as input; the top-scored *Mapped Term* for that input and the corresponding mapping score; a graphical depiction of the *neighborhood* of the ontology term, which contains the sub and superclasses of the term; a user option to specify the mapping type (Exact, Broad, or Narrow—subproperties of the standard *skos:mappingRelation* property defined in the SKOS vocabulary); and finally a dropdown menu for users to specify whether the mapping is approved or not.

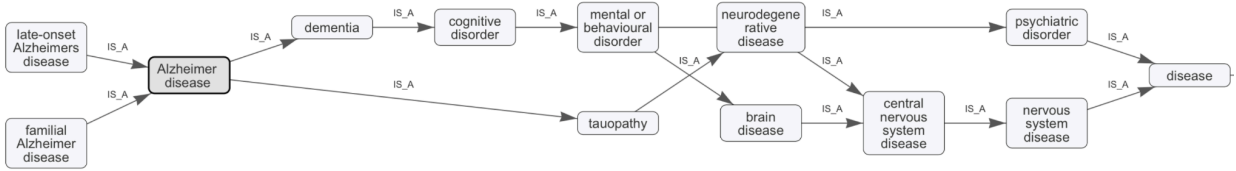

**Figure 4:** Ontology graph visualization provided by the text2term Web application. The visualization shows all ancestors of a term, in this case *Alzheimer disease*, as well as direct subclasses when they are less than 10. When a term has more subclasses, we represent them by a circle that can be expanded upon clicking.

## Usage Metrics

The text2term tool can be easily leveraged programmatically from a Python environment. After installation, there is a single function called *map_terms* that allows users to perform mapping. Since its inception in October 2022, the text2term Python package has been downloaded 17,025 times from PyPI.[2] On average, text2term is downloaded 946 times per month (see Figure 5)—since its first public release was on October 20, 2022, and the time of writing is April 15, 2024, we count these two half-months as one month. The text2term Python package has been installed via the *pip* installer a total of 829 times, and on average it is installed 46 times per month (Figure 5).

---

[2] We obtain download counts by querying the public PyPI download statistics dataset using Google BigQuery. Our queries are specified and documented in our evaluation repository for replicability: https://github.com/ccb-hms/ontology-mapper-evaluation.

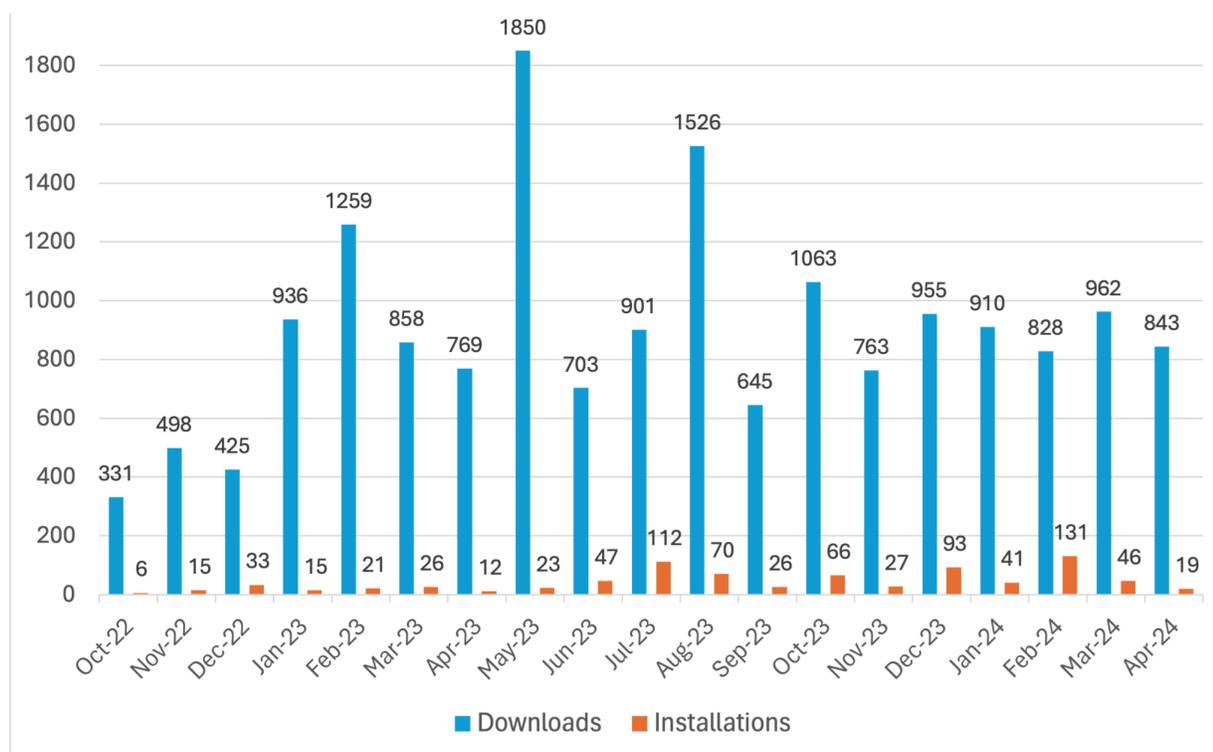

**Figure 5:** Number of downloads from PyPI and of installations via pip (*y*-axis) on each month since the first release on October 20, 2022, until the time of writing, April 15, 2024 (*x*-axis).

The text2term Web application hosted in our publicly accessible server has been lightly but regularly used since it first went live in August 2022. To date our server has processed just over 400 requests—on average, it receives about 20 requests per month.

## Comparison with Verified Mappings

In this section we describe a simple comparison between the mappings generated by our text2term tool and publicly available mappings verified by human curators. The initial motivation for the development of our tool was to facilitate bulk mapping of free-text descriptions of phenotypes to controlled terms in ontologies. As such, we attempted to identify ground-truth *benchmark* datasets that mimic precisely such a task of mapping phenotype descriptions to ontology terms.

### Methods and Materials

We identified a mapping set—the EFO-UKB mappings [11]—where the authors undertake the same task as we set out to do. We then add to our test corpus two generic collections of mappings between controlled terms in ontologies, controlled vocabularies, etc. Our test corpus consists of the three public mapping sets described below.

**EFO-UKB mappings**—a collection of mappings between phenotype descriptions in the UK Biobank and terms in the EFO ontology. Some of the phenotypes in the UK Biobank—a widely used source of population health data in research—are mapped to ICD-10. However, ICD-10 codes are not exhaustively mapped to public ontologies, which hinders interoperability of UK Biobank with public data. The authors built this mapping set using the Zooma ontology mapping tool followed by manual curation of mappings.

**OLS mappings**—a collection of mappings between biomedical ontologies that are hosted in the OLS repository [25]. These mappings are extracted directly from the ontologies, which have been specified by the respective ontology engineers. The OLS mappings are specified in the Simple Standard for Sharing Ontology Mappings (SSSOM) format [26].

**Biomappings**—a collection of community-contributed mappings between biomedical entities [10], which are not available from primary sources (such as OLS). The goal of this mapping collection is ultimately to be integrated with primary mapping sources and be distributed among other established mappings. Biomappings are also distributed in SSSOM format.

These mapping sets represent reasonable benchmarks and opportunities to assess the quality of our tool-generated mappings, since they have been verified by human experts and are useful public mapping sets that are either widely used or have the potential to be widely used.

Our comparison is limited to mappings to the EFO ontology, although the text2term tool is entirely generic and can be used with arbitrary OWL ontologies (or with ontologies hosted in Bioportal and OLS). We also limit our comparison to queries that have exactly one ontology mapping in the benchmark mapping set, since such a singular mapping is unambiguously indicative that one and only one term was the appropriate choice to annotate the given input. We use the TF-IDF-based mapper in text2term, as we observed that it is significantly faster and more effective than either edit-distance-based mappers and the Web API-based approaches.

We set out to compare these three mapping sets—our *benchmark mappings*—with the mappings generated by text2term for the same inputs, as depicted in Figure 6.

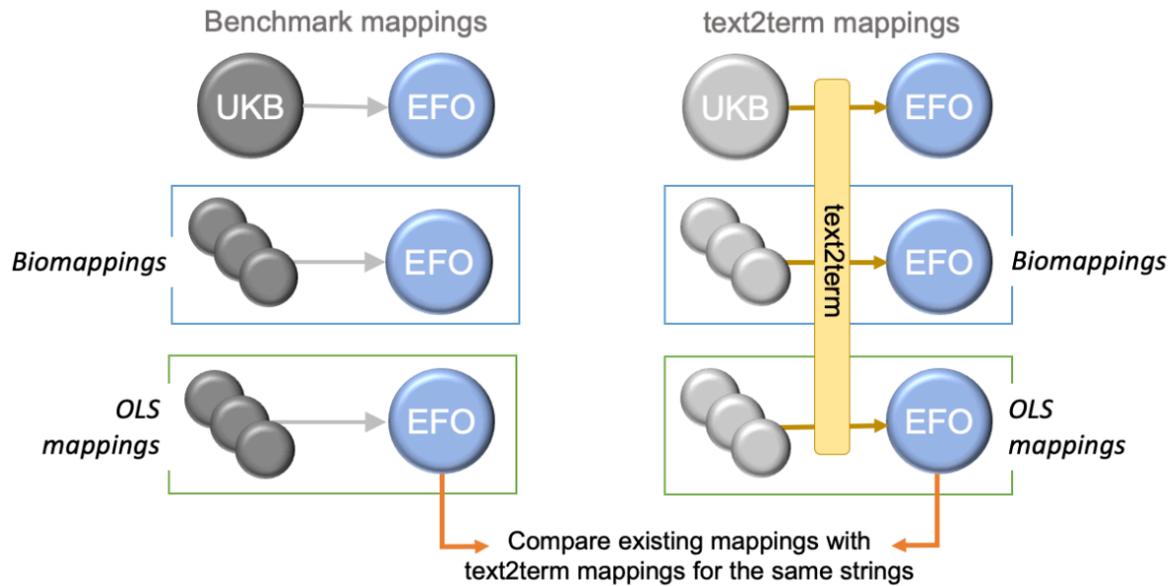

**Figure 6:** Overview of the design of our comparison between existing and text2term-generated mappings for the same text inputs. In the case of the Biomappings and OLS mappings, the text inputs are the labels of ontology terms that have been mapped to EFO.

Specifically, we want to determine whether text2term mappings coincide with the human-verified ones, whether they are more specific or more general (according to the ontology class hierarchy), whether they are siblings (i.e., they have a common direct parent), or whether they are (hierarchically) unrelated. In the case of the Biomappings and OLS mappings, we use the label of each *source* ontology term as the free-text input.

**Mapping categorization.** Consider a tool-generated mapping *T* and a human-verified mapping *H* for some input text *I*. We categorize *T* as:
- *Same:* if *T* is the same as *H*.
- *More specific*: if *T* is a subclass of *H*; i.e. EFO entails *T SubClassOf H*.
- *More general*: if *T* is a superclass of *H*; i.e, EFO entails *H SubClassOf T.*
- *Sibling*: if *T* and *H* have the same direct superclass; i.e. EFO states that *T SubClassOf P* and *H SubClassOf P*.
- *Unrelated*: if there is no sub or superclass relation between *T* and *H*.

We compute mappings of the (label) strings contained in each benchmark mapping set to the EFO ontology v3.62.0, released on January 15, 2024 (https://github.com/EBISPOT/efo/releases/tag/v3.62.0). We use text2term v4.1.2, released on March 6, 2024. All benchmark mapping sets and mappings computed by text2term in our comparison, along with the code to reproduce the entire comparison, are hosted in a GitHub repository, at https://github.com/ccb-hms/ontology-mapper-evaluation.

## Results

The results of the comparison of the mappings generated by text2term with those in our benchmark mapping sets are presented in Table 1.

| Category | UKB-EFO | Biomappings | OLS |
|---|---|---|---|
| *Same* | 660 (73%) | 626 (79%) | 6588 (81%) |
| *More Specific* | 34 (4%) | 2 (0.3%) | 91 (1.1%) |
| *More General* | 20 (4%) | 0 (0%) | 55 (0.7%) |
| *Sibling* | 13 (1%) | 47 (6%) | 89 (1.1%) |
| *Unrelated* | 172 (19%) | 120 (15%) | 1320 (16%) |

**Table 1: Results of comparison between mappings generated by text2term and mappings in our selected benchmark mapping sets.**

Overall, we observed that text2term finds (or re-finds) mappings with high accuracy in all our selected benchmark mapping sets. We think of this accuracy as a proxy for alleviated human effort, in the sense that, in a scenario where a curator is mapping a collection of strings, on average for some 3 out of 4 inputs, text2term picks the correct ontology mapping which the curator simply needs to confirm. In the remaining 1 out of 4 inputs, a curator may have to inspect additional potential mappings (remember, users can specify the number of mappings per input string) or resort to other resources.

We now look in more detail at the mappings where, for a given input entity, text2term did not find the expected benchmark mapping. We picked 3 examples in each category, and in each mapping set, to illustrate some patterns that we observe across many of the other mappings. The selected mapping pairs are shown in Tables 2-4.

| **Input Entity** | **text2term Mapping** | **Benchmark Mapping** | **Category** |
|---|---|---|---|
| *Granulomatous disorders of skin and subcutaneous tissue* | skin disease [EFO:000070] | granulomatous dermatitis [EFO:1000705] | More General |
| *hayfever or allergic rhinitis* | allergic rhinitis [EFO:0005854] | seasonal allergic rhinitis [EFO:0003956] | More General |
| *General pain for 3+ months* | pain [EFO:0003843] | Chronic pain [HP:0012532] | More General |
| *uremia* | uremia [EFO:1001226] | kidney failure [EFO:1002048] | More Specific |
| *thyroiditis* | thyroiditis [MONDO:0004126] | thyroid disease [EFO:1000627] | More Specific |
| *thyroid problem (not cancer)* | thyroid cancer | thyroid disease | More Specific |

|  | [MONDO:0002108] | [EFO:1000627] |  |
|---|---|---|---|
| *Gynecomastia* | Gynecomastia [HP:0000771] | breast hypertrophy [HP:0010313] | Sibling |
| *emphysema/chronic bronchitis* | chronic bronchitis [EFO:0006505] | emphysema [EFO:0000464] | Sibling |
| *bursitis* | bursitis [MONDO:0002471] | frozen shoulder [EFO:1000941] | Sibling |
| *Alzheimer s disease* | Alzheimer disease [MONDO:0004975] | Alzheimer's disease [EFO:0000249] | Unrelated |
| *Senile cataract* | senile cataract [MONDO:0004847] | age-related cataract [HP:0011141] | Unrelated |
| *Helicobacter pylori* | Helicobacter pylori infectious disease [EFO:1000961] | Helicobacter pylori [NCBITaxon:210] | Unrelated |

**Table 2: Example mappings selected from the results of the comparison of text2term with the UKB-EFO mappings.**

In Table 2, we see that some mappings selected by text2term are more appropriate for the input entities than the benchmark mappings— for example, "uremia", "thyroiditis", "Gynecomastia", "bursitis", "Senile cataract". Some other mappings provided by text2term can be considered just as correct as the benchmark mappings, for example in the cases of the mappings for the inputs "hayfever or allergic rhinitis", "emphysema/chronic bronchitis", or "General pain for 3+ months". In the Unrelated category, we observe that some benchmark mappings point to deprecated terms— for example, "Alzheimer's disease" [EFO:0000249] has been replaced (in EFO) by the term "Alzheimer disease" [MONDO:0004975], which text2term correctly identifies.

| **Input Entity** | **text2term Mapping** | **Benchmark Mapping** | **Category** |
|---|---|---|---|
| *Staphylococcal Infections [D013203]* | staphylococcal skin infections [EFO:1001849] | Staphylococcus aureus infection [EFO:0005681] | More General |
| *Venous Thrombosis [D020246]* | Venous thrombosis [HP:0004936] | deep vein thrombosis [EFO:0003907] | More General |
| *autoimmune disease [DOID:417]* | type II hypersensitivity reaction disease [EFO:0005809] | autoimmune disease [EFO:0005140] | Sibling |
| *Non-ST Elevated Myocardial Infarction [D000072658]* | ST Elevation Myocardial Infarction [EFO:0008585] | Non-ST Elevation Myocardial Infarction [EFO:0008586] | Sibling |
| *Calu-3 [CALU3_LUNG]* | Calu1 [EFO:0002151] | Calu3 [EFO:0002819] | Sibling |
| *childhood-onset asthma [DOID:0080815]* | childhood onset asthma [MONDO:0005405] | childhood onset asthma [EFO:0004591] | Unrelated |
| *Muscular Atrophy [D009133]* | Skeletal muscle atrophy [HP:0003202] | muscle atrophy [EFO:0009851] | Unrelated |
| *Neuralgia [D009437]* | neuralgia [EFO:0009430] | neuropathic pain [EFO:0005762] | Unrelated |

**Table 3: Example mappings selected from the results of the comparison of text2term with Biomappings.**

In Table 3, there are only 2 mappings considered more general than the ones in Biomappings—text2term correctly identifies more appropriate terms in both cases. In the 3 selected mappings (out of 47) in the Sibling category, the text2term mappings seem mostly incorrect when compared to Biomappings—this happens primarily because of numeric symbols or some form of negation in an input entity text (e.g. "non-"). In the Unrelated category, we start with another example of a benchmark mapping involving a deprecated term ('childhood onset asthma' [EFO:0004591]). In the other example mappings of the Unrelated category, the text2term recommendations seem just as reasonable as the benchmark mappings—and this highlights perhaps some repetition in EFO due to reuse of terms from other ontologies.

| Input Entity | text2term Mapping | Benchmark Mapping | Category |
|---|---|---|---|
| *Astrocytic Tumor [NCIT:C6958]* | astrocytic tumor [MONDO:0021636] | astrocytoma [EFO:0000272] | More General |
| *meningitis [DOID:9471]* | meningitis [MONDO:0021108] | infectious meningitis [EFO:0000584] | More General |
| *Carcinoma de tiroides [HP:0002890]* | carcinoma [EFO:0000313] | thyroid carcinoma [EFO:0002892] | More General |
| *cellular schwannoma [DOID:3196]* | cellular schwannoma [MONDO:0002548] | schwannoma [EFO:0000693] | More Specific |
| *brain cancer [DOID:1319]* | brain cancer [MONDO:0001657] | brain neoplasm [EFO:0003833] | More Specific |
| *Dandy-Walker Malformation [NCIT:C75012]* | isolated Dandy-Walker malformation with hydrocephalus [MONDO:0017110] | Dandy-Walker syndrome [EFO:1000890] | More Specific |
| *stroke disorder [MONDO:0005098]* | stroke disorder [MONDO:0005098] | stroke [EFO:0000712] | Sibling |
| *melancholic depression [DOID:1595]* | melancholia [EFO:1002014] | unipolar depression [EFO:0003761] | Sibling |
| *polyarteritis nodosa [DOID:9810]* | polyarteritis nodosa [MONDO:0019170] | Polyarteritis Nodosa [EFO:0009012] | Sibling |
| *vein [XAO:0000115]* | vein disorder [MONDO:0004634] | obsolete_vein [EFO:0000816] | Unrelated |
| *Kaposi's sarcoma cell* | Kaposi's sarcoma | Kaposi's sarcoma cell | Unrelated |

| [BTO:0002071] | [EFO:0000558] | [EFO:0000187] | |
| *tuberculosis* [DOID:399] | tuberculosis [MONDO:0018076] | obsolete_tuberculosis [EFO:0000774] | Unrelated |

**Table 4: Example mappings selected from the results of the comparison of text2term with the OLS mappings.**

In Table 4, we see once again examples of mappings, both in the More General and the More Specific categories, where text2term identified just as correct terms compared to the benchmark mappings (e.g., "Astrocytic tumor", "brain cancer"), some times more appropriate terms (e.g., "meningitis", "cellular schwannoma" and "stroke disorder"), and other times not as suitable terms (e.g., "Carcinoma de tiroides", "Dandy-Walker Malformation"). In the Sibling category, text2term tends to find more appropriate terms than those in the benchmark mappings—e.g., "stroke disorder, and "melancholic depression", which is more appropriately mapped to melancholia than to unipolar depression. The last example mappings for 'polyarteritis nodosa' reveal again some conceptual duplication within EFO, which contains two non-deprecated terms that unambiguously are intended to describe the same condition. Finally, most of the mappings in the Unrelated category in the OLS mappings are related to obsolete terms in EFO, some of which have been replaced by terms that text2term correctly maps the input entity to—for example, "vein" and "tuberculosis".

## Discussion and Conclusions

The text2term tool is a simple yet versatile open-source tool for mapping free-text descriptions of biomedical entities to controlled terms in ontologies. It is easy to use and can be leveraged in multiple ways to cater to different users and user backgrounds—text2term can be used programmatically as a Python package; via a command-line interface; via our Harvard-hosted, graphical user interface-based Web application (https://text2term.hms.harvard.edu); or by deploying a local instance of the Web application using our provided Docker image. The tool is versioned and documented on GitHub, where we also keep a public issue tracker.

The text2term tool is developed by the Center for Computational Biomedicine at the Harvard Medical School and is actively developed. It is used both internally for various projects that build data assets for the biomedical community and by a growing external user community. The usage metrics we collected are suggestive of steady demand and usage of our tool. A key design principle behind text2term is to have an extendable collection of mappers, which can be leveraged by users to plug-in a different mapping algorithm while taking advantage of the tool's remaining infrastructure and user interfaces to browse and verify the resulting mappings. By default, text2term supports string comparisons using edit distance-based algorithms such as Levenshtein distance; a TF-IDF-based algorithm that compares strings based on ngrams of configurable size; the BioPortal Annotator Web service, which makes it possible to map to any ontology available in the BioPortal ontology repository; or the Zooma annotator that enables mapping to ontologies in the Ontology Lookup Service (OLS) repository.

Our evaluation of text2term demonstrated that the tool is highly effective at finding appropriate ontology terms to represent input free-text descriptions of biomedical entities. This finding suggests that using such a tool for bulk standardization of free-text entity descriptions can substantially alleviate the amount of human curation needed, since for most inputs a curator only needs to confirm the mappings generated by the tool. When compared to the (cross-ontology) mappings that are encoded in OLS ontologies or in Biomappings, which are contributed and/or reviewed by domain experts, text2term actually exposed some wrong mappings in the current versions of the benchmark mappings sets, as well as some modeling issues in the EFO ontology itself—particularly the existence of duplicate terms to represent the same concept.

The potential applications of text2term are twofold. First, it can be used to retrospectively standardize free-text entity descriptions in existing data with ontologies. This can be done programmatically using the highly customizable Python package or the command-line interface, or interactively using the text2term Web application that can be deployed using Docker. Second, text2term can be used to prospectively control the input of terms as part of an auto-complete mechanism—for example in a (Web) form—that checks user input against ontology terms, and displays the most appropriate term options. In this way, all inputs to specific fields that should be grounded with ontologies (for example, disease or organism names, cell types, etc.), are guaranteed to be controlled, resolvable terms from one or more ontologies of choice.

In future work, we plan to continue to build and maintain text2term based on requirements from projects at the Harvard Medical School, as well as based on public requests for features and enhancements. For example, while we initially built text2term to map strings to ontology classes, a group of users was interested in mapping to ontology properties in order to standardize the names of relationships between entities in some data. So we enhanced the tool to allow users to specify whether they wish to map to ontology classes, properties, or both. An additional future work direction for us is to more precisely map terms that include some form of negation—in its current form, the tool favors, for example, an ontology term such as "hodgkin's lymphoma" for an input string such as "non-hodgkin's lymphoma", potentially because "non" is considered a stop word (and thus removed) by one of the libraries that we reuse.

## Conflict of interest

Robert Gentleman consults broadly in the Biotech industry and owns shares or options in many companies.

## Acknowledgements

We thank our collaborators from the Harvey Mudd College: Carmen Benitez, Cindy Lay, An Nguyen, Kobe Rico, and Matthew Waddell, under the supervision of Jamie Haddock, for their contributions to the design and development of text2term.

## Data availability

The text2term tool is freely available, open-source and distributed under the MIT License. The backend Python package is hosted and versioned on GitHub (https://github.com/ccb-hms/ontology-mapper), and distributed via PyPI (https://pypi.org/project/text2term). The frontend Web application codebase is also hosted and versioned on GitHub (https://github.com/ccb-hms/ontology-mapper-ui). The application can be deployed in local systems using Docker, following the instructions provided in the GitHub repository. A fully functional instance of the Web application is available for public use in our server at https://text2term.hms.harvard.edu. For large jobs or to avoid network latency, we recommend using the Docker installation or the Python package.

The code to reproduce the evaluation and the resulting mapping comparison data can be found at: https://github.com/ccb-hms/ontology-mapper-evaluation.

# Supplementary Material

## Configuration Options

| Option name | Type | Description |
|---|---|---|
| **source_terms*** | string/list/ dictionary | One of the following: 1) a string that specifies a path to a file containing the terms to be mapped, 2) a list of the terms to be mapped, or 3) a dictionary where each key is a term to be mapped, and each value is a list of tags |
| **target_ontology*** | string | Path or URL or acronym of 'target' ontology to map the source terms to. When the chosen mapper is BioPortal or Zooma, provide a comma-separated list of ontology acronyms (eg 'EFO,HPO') or write 'all' to search all ontologies. When the target ontology has been previously cached, provide the ontology name that was used to cache it. As of version 2.3.0, it is possible to specify ontology acronyms as the target_ontology (eg "EFO" or "CL"), which is achieved using bioregistry to retrieve URLs for those acronyms |
| **base_iris** | tuple | Map only to ontology terms whose IRIs start with one of the strings given in a tuple, for example: ('http://www.ebi.ac.uk/efo','http://purl.obolibrary.org/obo/HP') |
| **csv_column** | tuple | Specify a column to map if a CSV is the input file. Ignored otherwise. |
| **source_terms_ids** | tuple | Collection of identifiers for the given source terms |
| **excl_deprecated** | boolean | Exclude ontology terms stated as deprecated via *owl:deprecated true* |
| **mapper** | Mapper | Method used to compare source terms with ontology terms. One of: levenshtein, jaro, jarowinkler, jaccard, indel, tfidf, zooma, bioportal |
| **max_mappings** | int | Maximum number of mappings returned per source term |
| **min_score** | float | Minimum similarity score [0,1] for the mappings (1=exact match) |
| **output_file** | string | Path to the output file for mappings |
| **save_graphs** | boolean | Save graphs representing the neighborhood of each ontology term |
| **save_mappings** | boolean | Save the generated mappings to a file (specified by output_file) |
| **separator** | string | Character that separates source term values in a character-separated file input. Ignored otherwise |
| **use_cache** | boolean | Use a previously cached ontology |
| **term_type** | Ontology TermType | Specifies whether to map to ontology classes, properties, or both |
| **incl_unmapped** | boolean | Include all unmapped terms in the output, even if they are scored below the min_score |

**Supplementary Table S1: Configuration options available in text2term**. The first two arguments, marked with an asterisk (*), are the only required arguments; all other arguments are optional. The options are available programmatically and through a command-line interface.

## Additional Tool Details

Here we describe additional details about the design, usage, and testing of the text2term tool.

**Input terms.** When working with tabular input, users can specify the table column containing values to be standardized via an argument '*csv_column*', while another argument 'source_term_ids' can be used to specify what column (if any) contains identifiers associated with the source terms. When used programmatically, the input can take two additional forms: it can be a list of terms or a dictionary where each key is a term to be mapped and each value is a list of tags associated with the term during preprocessing.

**Input terms preprocessing.** text2term includes regular expression-based preprocessing functionality for input terms. The preprocessing function takes the input terms and a collection of user-defined regular expressions, and then matches each term to each regular expression to simplify the input term text. This is particularly useful when there are patterns of noisiness in the terms, whereby extra words in the term are either not necessary for the mapping, or they are contributing to poor mappings. Additionally, the tool can be configured to ignore user-defined patterns of strings, specified via regular expressions in a "blocklist" file. Blocklisted terms are not mapped to the ontology but are included in the output with a tag denoting they were ignored.

**Exclude deprecated terms.** The text2term tool can be configured to exclude deprecated ontology terms from the potential term matches, which are those terms with an OWL assertion of the form *owl:deprecated true*. This is desirable to avoid using terms that are potentially not maintained in future ontology versions.

**Filter ontology terms by their IRIs.** It is common that ontologies reuse terms from external ontologies, rather than creating new terms for entities that have been formalized elsewhere. For example, the MONDO disease ontology [22] contains relationships between diseases and anatomical locations, where the anatomical locations are terms from the Uberon anatomy ontology. To make the mapping process more efficient in text2term, it is possible to target the mapping to terms whose identifiers are within the namespaces of specific ontologies. For example, to map only to disease terms from MONDO, a user would specify a list of allowed namespaces containing the base IRI of MONDO terms (http://purl.obolibrary.org/obo/MONDO).

**Ontology cache for faster mapping.** We implemented a caching mechanism that allows users to cache ontologies that they intend to use more than once. A user can cache ontologies using either the *cache_ontology* function to cache a single ontology from a URL or local file path; or the *cache_ontology_set* function to cache a collection of ontologies specified in a table, also by their URLs or local file paths. The first of these will cache a single ontology, with it being identified by an acronym that will be used to reference it later. Once an ontology has been cached by either function, it is stored in a local cache folder and can be accessed by using the assigned acronym in the *target_ontology* (e.g., "EFO") and setting the *use_cache* flag to *True*.

**Options to filter results.** The text2term tool provides various filtering options (see Table S1). It is possible to specify the maximum number of mappings to compute for each input term

('max_mappings'). Also, users can specify the minimum "acceptable" score for their mappings ('min_score'), in a scale from 0 to 1, where 1 implies an exact match. One can also choose to map to either or both class and property term types, which are the two typical ontology term types used in biomedical ontologies.

**Continuous integration/testing:** We developed a comprehensive unit and integration test suite that is triggered by a GitHub Action every time there is a Git commit to the *main* branch. We employ the best practice of developing against a *development* Git branch, and only committing to *main* code that has been locally tested and reviewed. Then by pushing development changes to the main branch, our GitHub Action triggers the execution of the test suite, which then reports whether all tests passed, or which tests in our test suite failed and why.